\documentstyle[multicol,aps,psfig]{revtex}
\begin{document}

\title{ 
Squeezed Correlations and Spectra for Mass-Shifted Bosons }

\author{M. Asakawa$^1$, T. Cs\"org\H o$^{2,3}$, and M. Gyulassy$^{3}$ }

\address{$^1$ Dept. Physics, School of Science,\\
Nagoya University, Nagoya, 464 - 8602, Japan\\
$^2$ MTA KFKI RMKI, H - 1525 Budapest 114, POB. 49, Hungary \\
$^3$  Dept. Physics, Columbia University, \\
        538 W 120-th Street, New York, NY 10027, USA}

\maketitle

\begin{abstract}
Huge back-to-back correlations are shown to arise
for thermal ensembles of bosonic states 
with medium-modified masses. The effect is experimentally 
observable in high energy heavy ion collisions.
\end{abstract}
 
\begin{multicols}{2} 
{\it Introduction} --- 
The Hanbury Brown - Twiss (HBT) or Goldhaber - Goldhaber - Lee - Pais
(GGLP) effect ~\cite{hbt,gglp} 
is  widely used in heavy ion physics to measure the space-time geometry
of such reactions. The enhanced correlations
of bosons in outgoing states with small relative momentum
provide a Fourier transformed picture
of the system at freeze-out. The scales measured
by the HBT effect coincide with the lengths of homogeneity~\cite{sm}. 

In this Letter we consider the effect of possible
mass shifts in the dense medium on two boson correlations in general.
Thus far medium modifications of hadron masses have been mainly
considered in terms of effects on such observables as
dilepton yields and spectra.
Hadron mass shifts are caused by interactions
in a dense medium and therefore vanish on the freeze-out surface.
Thus, a naive first expectation is that in-medium hadron-modifications 
may have little or no effect on two boson correlations,
and so the usual HBT effect has been expected to be only concerned
with the geometry and matter flow gradients on the freeze-out surface.
However, in this Letter we show that an interesting
quantum mechanical correlation is induced due to the
fact that medium modified bosons can be represented in terms of
two-mode squeezed states of the asymptotic bosons,
which are observables.
As a by-product, we solve the finite-size problem for
two-mode squeezing, 
by formulating the theory of intensity interferometry
for a finite, inhomogeneous, squeezed and expanding medium,
which is a new result also for 
quantum optics~\cite{janszky}. 

In this Letter we assume the validity of relativistic
hydrodynamics up to freeze-out.
The local temperature $T(x)$, chemical potential $\mu(x)$,
and flow fields $u^{\mu}(x)$, are given, 
for example, by those taken from ref.~\cite{rg}.
In relativistic heavy ion collisions, it has been observed
that the one particle spectra can be described by thermal
distribution fairly well \cite{na44}.
We assume that the sudden (non-adiabatic) approximation
is a valid abstraction in describing
the freeze-out process in relativistic heavy ion collisions
quantum mechanically 
\cite{aw}
and that there exists an abrupt freeze-out
surface, $\Sigma^\mu(x)$. However, the effects of 
fluctuations of that freeze-out process
will also be considered below.
        
Consider the following model Hamiltonian, 
\begin{eqnarray}
{H} & = &  H_0 - \frac{1}{2} \int d {\bf x} d {\bf y} \phi({\bf x})
\delta M^2({\bf x}-{\bf y}) \phi({\bf y}), 
\label{ham} \\
        H_0 & = & \frac{1}{2} \int d {\bf x} \left(
                \dot{\phi}^2+ |\nabla \phi|^2
                +
                        m_0^2 \phi^2  \right),
\end{eqnarray}
where $H_0$ is
the asymptotic Hamiltonian, in the rest frame of matter.
The scalar field $\phi({\bf x})$ in the Hamiltonian $H$ corresponds to 
quasi - particles that propagate with a momentum-dependent
medium-modified effective mass,
which is related to the vacuum  mass, $m_0$,  via
$$
 m_*^2({|{\bf k}|}) =  m_0^2 - \delta M^2({|{\bf k}|}).
$$
The mass-shift is assumed to be limited to long wavelength 
collective modes :
$
\delta M^2({|{\bf k}|}) \ll m_0^2$ if $|{\bf k}| > \Lambda_s.$ 

We are interested in the invariant
single-particle and two-particle momentum distributions:
\begin{eqnarray}
N_1({\bf k}_1) & = & \omega_{{\bf k}_1}{d^3N \over d{\bf k}_1} 
        = \omega_{{\bf k}_1} \langle
 a^\dagger_{{\bf k}_1} a^{\phantom\dagger}_{{\bf k}_1}\rangle , \\
N_2({\bf k}_1,{\bf k}_2) & = & 
\omega_{{\bf k}_1} \omega_{{\bf k}_2} 
        \langle a^\dagger_{{\bf k}_1} a^\dagger_{{\bf k}_2} 
        a^{\phantom\dagger}_{{\bf k}_2}
        a^{\phantom\dagger}_{{\bf k}_1} \rangle ,\\
\langle a^\dagger_{{\bf k}_1} a^\dagger_{{\bf k}_2} 
a^{\phantom\dagger}_{{\bf k}_2} a^{\phantom\dagger}_{{\bf k}_1} \rangle 
& = &  
\langle a^\dagger_{{\bf k}_1} a^{\phantom\dagger}_{{\bf k}_1}\rangle
\langle  a^\dagger_{{\bf k}_2} a^{\phantom\dagger}_{{\bf k}_2} \rangle  + 
        \nonumber \\
        \null &  &  
\hspace*{-1.cm}+\,
\langle a^\dagger_{{\bf k}_1} a^{\phantom\dagger}_{{\bf k}_2}\rangle
\langle  a^\dagger_{{\bf k}_2} a^{\phantom\dagger}_{{\bf k}_1} \rangle + 
\langle a^\dagger_{{\bf k}_1} a^\dagger_{{\bf k}_2}\rangle
\langle  a^{\phantom\dagger}_{{\bf k}_2}
         a^{\phantom\dagger}_{{\bf k}_1} \rangle,
\label{rand}
\end{eqnarray}
where $a_{\bf k}$ is the annihilation operator for the
asymptotic quantum with four-momentum 
$k^{\mu}\, = \, (\omega_{\bf k},{\bf k})$,
$\omega_{\bf k}^2= m_0^2 + {\bf k}^2$ ($\omega_{\bf k} > 0$)  and
the expectation value of an operator $\hat{O}$ is given with
the density matrix $\hat{\rho}$ as 
$\langle \hat{O} \rangle = {\rm Tr} \, \hat{\rho}\, \hat{O}$.
Eq.(\ref{rand})
has been derived as a generalization
of Wick's theorem for {\em locally} equilibrated (chaotic) 
systems in refs.\cite{sm,gykw,surp}.

In order to simplify notation, we introduce the 
chaotic and squeezed amplitudes, defined, respectively, as
\begin{eqnarray}
G_c(1,2) & = & 
\sqrt{\omega_{{\bf k}_1} \omega_{{\bf k}_2} }
  \langle a^{\dagger}_{{\bf k}_1} a^{\phantom\dagger}_{{\bf k}_2}\rangle,
\label{gc}\\
G_s(1,2) & = & 
\sqrt{\omega_{{\bf k}_1} \omega_{{\bf k}_2} }
  \langle a^{\phantom\dagger}_{{\bf k}_1}
  a^{\phantom\dagger}_{{\bf k}_2} \rangle .
\label{gs}
\end{eqnarray}
Usually, the chaotic amplitude, $G_c(1,2) \equiv G(1,2)$
is dominant, and carries the Bose-Einstein correlations,
while the squeezed amplitude, $G_s(1,2)$ vanishes:
\begin{equation}
C_2({\bf k}_1,{\bf k}_2)  =  {N_2({\bf k}_1,{\bf k}_2)
\over N_1({\bf k}_1) N_1({\bf k}_2) } = 
        1 + { | G(1,2) |^2 \over G(1,1) G(2,2) }.
\label{hbt}
\end{equation}
The exact value of the intercept, $C_2({\bf k},{\bf k})=2$,
is a characteristic signature  of a chaotic Bose gas
without dynamical 2-body correlations outside the domain
of Bose-Einstein condensation~\cite{cstjz}.

For a hydrodynamic ensemble, 
eq. (\ref{gc})
reduces to the special form derived by Makhlin and Sinyukov~\cite{sm},
\begin{equation}
G(1,2) = { \displaystyle\phantom{|}
{1 \over\displaystyle\phantom{|}
(2\pi)^3
}} \,
\int  d^3 \Sigma_\mu K^\mu_{1,2} e^{i q_{1,2}\cdot x}\; n(x,K_{1,2})
\; \; .\label{num}
\end{equation}
Note that the relative and average pair
momentum coordinates, 
$q^0_{1,2}=\omega_1-\omega_2$,
${\bf q}_{1,2}={\bf k}_1-{\bf k}_2$, and 
${\bf K}_{1,2}={\textstyle\frac{1}{2}}({\bf k}_1+{\bf k}_2)$
appear in
(\ref{num}). 
These variables arise naturally whenever  the Wigner operator is used,
even in totally non-equilibrium semi-classical limits\cite{pgg}. 
The validity of the approximations leading to
(\ref{num}) 
requires  the width of $G(1,2)$ as a function
of the relative momentum, $q = |{\bf q}_{1,2}|$, to be small. That width 
is given by $\sim 1/R$, where $R$ is
a characteristic dimension of
the system. The semi-classical limit corresponds
to $KR\gg 1$, where $K$ is $|{\bf K}_{1,2}|$.
Note that $\sqrt{\omega_{k_1}  \omega_{k_2}}\sim K^0_{1,2}$
in this case.
For $qR< 1$, the second term in (\ref{rand})
describes the minimal quantum interference
associated with the indistinguishability of the bosons.
The integration over
the freeze-out surface, $\Sigma^\mu(x)$, 
is implemented with the invariant measure
$d^3 \Sigma_\mu K^\mu_{1,2}$
that reduces to $K^0 d^3x$ in the special case of 
a constant freeze-out time.

{\it Results for a homogeneous system} --
The terms neglected in (\ref{hbt}) involving $G_s(1,2)$
become non-negligible when mass shift becomes non-vanishing,
i.e., $\delta M^2(|{\bf k}|)\ne 0$. Given such a mass shift,
the dispersion relation is modified to
$\Omega_{\bf k}^2 =\omega^2_{\bf k}-\delta M^2(|{\bf k}|)$,
where $\Omega_{\bf k}$ is the frequency of the in-medium mode
with momentum ${\bf k}$.
The annihilation operator for the in-medium quasi-particle with
momentum ${\bf k}$, $b_{\bf k}$, and that of the asymptotic
field, $a_{\bf k}$, are related by a Bogoliubov
transformation \cite{ac}: 
\begin{equation}
a^{\phantom{\dagger}}_{{\bf k}_1}
        = c^{\phantom{\dagger}}_{{\bf k}_1}
          b^{\phantom{\dagger}}_{{\bf k}_1} + 
                s^{*\phantom{\dagger}}_{-{\bf k}_1} b^\dagger_{-{\bf k}_1}
\equiv C^{\phantom{\dagger}}_1 + S^\dagger_{-1},
\label{asq}
\end{equation}
where
$c_{\bf k}=\cosh[r_{\bf k}]$, 
$s_{\bf k}=\sinh[r_{\bf k}]$  and $r_{\bf k}$
is given by
\begin{equation}
r_{\bf k}=\frac{1}{2}\log(\omega_{\bf k}/\Omega_{\bf k})\;\; .\label{sqr}
\end{equation}
We introduce the shorthand, $C^{\phantom{\dagger}}_1$ and $S_{-1}^\dagger$,
to simplify later notation. As is well-known, the Bogoliubov
transformation is equivalent to a squeezing operation, and so
we call $r_{\bf k}$ the mode dependent squeezing parameter.
While it is the $a$-quanta that are observed, it is the $b$-quanta
that are thermalized in medium. Thus, we consider the
thermal average for a globally thermalized gas of 
the $b$-quanta, that is homogeneous in volume $V$:
\begin{equation}
\hat{\rho} = {\displaystyle\phantom{|} 1 \over Z} \exp\left(-\frac{1}{T}
        \frac{V}{(2 \pi)^3} \int d {\bf k}\, \Omega_{\bf  k}
                \, b^\dagger_{\bf k} b^{\phantom\dagger}_{\bf k}\right).
\end{equation}
When this thermal average is applied,
\begin{eqnarray}
G_c(1,2) & = &
\sqrt{\omega_{{\bf k}_1} \omega_{{\bf k}_2}}
\left[\langle C^\dagger_1C^{\phantom\dagger}_2\rangle + 
\langle
S^{\phantom\dagger}_{-1} S^\dagger_{-2}\rangle\right], \label{gc1}\\
G_s(1,2)&=&
\sqrt{\omega_{{\bf k}_1} \omega_{{\bf k}_2}}
\left[
\langle S^\dagger_{-1} C^{\phantom\dagger}_2
\rangle + \langle C^{\phantom\dagger}_1 S^\dagger_{-2}
\rangle 
\right]
. \label{gs1}
\end{eqnarray}

If this thermal $b$ gas freezes out suddenly at some time at
temperature $T$,
the observed single $a$-particle distribution takes the following form:
\begin{eqnarray}
N_1({\bf k}) & = & \frac{V}{(2 \pi)^3}\, \omega_{\bf k}\, n_1({\bf k}),\\
n_1({\bf k}) & = &
        | c_{\bf k}^{\phantom{\dagger}}|^2 
        n_{\bf k}^{\phantom{\dagger}} 
        + |  s_{-\bf k}|^2 (n_{-\bf k} + 1), \\\
n_{\bf k} & = & \frac{1}{\exp(\Omega_{\bf k}/T) -1} \; .
\end{eqnarray}
This spectrum includes a squeezed vacuum contribution
in addition to the mass modified thermal spectrum. 
Its shape, 
however, remains essentially that of the conventional thermal distribution.
Although the squeezed vacuum contribution results in a slowly decaying
power-law tail of the single-particle spectra,
this power-law tail arises
for typical values of mass-shifts and temperatures 
at high transverse momentum only, e.g. above $|{\bf k}| > 1$ GeV,
where the local hydrodynamic description breaks down  and correlated 
pQCD (mini-jets)  starts to play a dominant role. 
We emphasize that the spectra and correlation functions
should be described by the same set of model parameters, and 
any signal found in the correlation function could be cross-checked
against its contribution to $N_1({\bf k}) $.

In the homogeneous limiting case, $G_c(1,2)\propto V \delta_{1,2}$,
while $G_s(1,2)\propto V \delta_{1,-2}$.
The resulting two particle correlation function is
therefore unity except for the parallel and antiparallel cases: 
\begin{eqnarray}
C_2({\bf k}, {\bf k}) & = & 2, \label{e:c2}\\
C_2({\bf k}, {\bf -k}) & = & 1 +
{\displaystyle\phantom{|} {|c^*_{\bf k}s_{\bf k}^{\phantom{\dagger}}
        n_{\bf k}^{\phantom{\dagger}}+
        c^*_{-\bf k} s_{-\bf k}^{\phantom{\dagger}}
        (n_{-\bf k}^{\phantom{\dagger}}  + 1) |^2 }
\over \displaystyle\phantom{|} {n_1({\bf k}) \, n_1({- \bf k})} }. 
        \label{e:c2b}
\end{eqnarray}
The {\em dynamical} correlation due to the two mode squeezing associated
with mass shifts is therefore
{\em back-to-back} as first pointed out in \cite{ac}.
The HBT correlation intercept 
remains 2 for identical momenta. Evaluating eq.~(\ref{e:c2b}) for $T = 140$
MeV, $|{\bf k}| = 0$, 300 or 500 MeV for $\phi$ mesons,
as a function of the medium modified $m^*_{\phi}$, one finds 
back-to-back correlations (BBC) as big as 100 - 1000
for reasonable values of $m^*_{\phi}$. 

It follows from eq.~(\ref{e:c2b}) that the BBC
are unbounded from above, $1 \leq C_2({\bf k},-{\bf k}) < \infty $.
As $|{\bf k}| \rightarrow \infty$, 
$C_2({\bf k}, {\bf -k}) \simeq
1 + 1 / |s_{\bf -k}|^2 \simeq 1 + 1/n_1({\bf k}) \rightarrow \infty$.
Hence at large values of ${\bf k}$, particle production
is {\it dominated} by that of back-to-back correlated pairs for any
non-vanishing value of 
in-medium mass-shifts. This huge enhancement
renders our effect measurable. This prediction has to be contrasted to the
results of ref.~\cite{aw}, which predicted large values for the
HBT correlations, $C({\bf k}, {\bf k})$ as a consequence of an
incorrect Bogoliubov transformation. Ref.~\cite{surp} discussed a different kind
of back-to-back correlation, the particle - antiparticle correlation (PAC),
which is based on a classical current formalism,
and showed 
$C_2^{(PAC)}({\bf k}, {\bf -k}) \leq 3$.
However, their upper limit was shown to be reachable for massless quanta
at $|{\bf k}| = 0$ only. For massive final state bosons and for any 
realistic finite duration of particle emission, the effects
found in ref.~\cite{surp} are suppressed by at least 2 -3 orders
of magnitude, resulting in an unmeasurably small effect.
In contrast, we consider here an entirely quantum effect,
 which cannot be obtained in classical current formalism. 
The {\it unlimited strength} of the back-to-back correlations 
of mass-shifted mesons in our case results in a {\it measurable effect}
even if the decay of the medium is not completely sudden
and orders of magnitude suppressions reduce its strength.

{\it Suppression by finite emission times}.
To describe a more gradual freeze-out, the 
probability distribution $F(t_i)$ of the decay times $t_i$
is introduced. (The sudden approximation is recovered in the 
$F(t_i)= \delta(t_i - t_0) $ limiting case.)
The time evolution of $a_{\bf k}(t)$
will be $a_{\bf k}(t) = a_{\bf k}(t_i) \exp[-i \omega_{\bf k} (t - t_i)]$,
which leads to
\begin{eqnarray}
C_2({\bf k}, {\bf -k}) & - & 1 \,\,  = \nonumber \\
&&\hspace*{-1.8cm} = 
{\displaystyle\phantom{|} {|c^*_{\bf k}s_{\bf k}^{\phantom{\dagger}}
        n_{\bf k}^{\phantom{\dagger}}+
        c^*_{-\bf k} s_{-\bf k}^{\phantom{\dagger}}
        (n_{-\bf k}^{\phantom{\dagger}}  + 1) |^2 }
\over \displaystyle\phantom{|} {n_1({\bf k}) \, n_1({- \bf k})} }
        |\tilde F(\omega_{\bf k} + \omega_{-{\bf k}}) |^2 , 
         \label{e:c2bs}
\end{eqnarray}
where $\tilde F(\omega) = \int dt F(t) \exp(-i \omega t)$.
For a typical exponential decay, 
$F(t)  = \theta(t-t_0)\Gamma \exp[-\Gamma (t-t_0)] $, 
the suppression factor is 
$$
|\tilde F(\omega_{\bf k} + \omega_{-{\bf k}}) |^2 = 
1/[ 1 + (\omega_{\bf k} + \omega_{-\bf k})^2 / \Gamma^2].  
$$
In the adiabatic limit, $\Gamma\rightarrow 0$, this factor
suppresses completely the BBC, while in the sudden approximation,
$\Gamma\rightarrow \infty$, the full BBC are preserved.
For $ \delta t = \hbar/\Gamma = 2$ fm/c,  
and for BBC of $\phi$ mesons with  $\omega \sim 1-2$ GeV, 
one finds that the BBC are suppressed by factor $\sim 10^{-3}$.
As shown in Fig. 1, the BBC survive this large
suppression with a measurable strength,
 as large as 2-3, the scale of HBT correlations.
This emphasizes the enormous strength of squeezed BBC for mass-shifted bosons.

{\it Results for inhomogeneous systems}.
Following Ref.\cite{sm}, we divide the inhomogeneous fluid into
independent cells labeled $i$ and assume
\begin{eqnarray}
\hat{\rho} & = & \prod_{i} \hat{\rho}_i \, = \,
        \prod_i {\displaystyle\phantom{|} 1  \over\displaystyle\phantom{|} Z_i}
\exp\left[ - (H_i - \mu_i \tilde{N}_i) / T_i \right],
\end{eqnarray}
where $\hat{\rho}_i$ is the local thermal density matrix
of cell $i$ at the decoupling time $t_i$, and $\tilde{N}_i$ is the local number
operator of $b$
quanta.  
The hydrodynamic limit applies
only if the cells are small compared to the scale of change of the 
temperature, flow and chemical potential fields, however, they are big enough
so that one can apply quantization and statistics cell by cell. 
The validity of hydrodynamics in heavy ion collisions
is not obvious, but must be checked in each reaction
through the consistency with both single and double inclusive
spectra.
In each cell, the field can be expanded with 
creation and annihilation operators, 
and $H_i$ is diagonalized by a local
Bogoliubov transformation.
The amplitudes (\ref{gc1},\ref{gs1}) can be evaluated
assuming that the $b$-quanta satisfy locally  a generalized eq.~(\ref{asq}):
\begin{eqnarray}
G_c(1,2) & = & 
        {\displaystyle\phantom{|} 1 \over (2 \pi)^3 }
        \int d^4\sigma_{\mu}(x) K_{1,2}^{\mu} e^{i q_{1,2} \cdot x} \nonumber \\
        && \times 
         \left[|c_{1,2}|^2   n_{1,2} + 
                | s_{-1,-2}|^2  (n_{-1,-2} + 1) \right],\label{e:gc} \\
G_s(1,2) & = & 
        {\displaystyle\phantom{|} 1 \over (2 \pi)^3 }
        \int d^4\sigma_{\mu}(x) K_{1,2}^{\mu} e^{2 i  K_{1,2} \cdot x}  
	\nonumber \\
        && \hspace{-0.6cm} \times \left[
                s^*_{-1,2} c_{2,-1} n_{-1,2}  +
                c_{1,-2} s^*_{-2,1} (n_{1,-2} + 1)
                \right]. 
        \label{e:gd}\label{e:gs}
\end{eqnarray}
Here $d^4\sigma^{\mu}(x) = d^3\Sigma^{\mu}(x;\tau_f)\, F(\tau_f)  
d\tau_f $ is the product of the normal-oriented
volume element depending parametrically on $\tau_f$ (the freeze-out 
hypersurface parameter) and the
invariant distribution of that parameter $F(\tau_f)$, 
and $u^\mu(x)$ is the local flow vector at freeze-out.
The other variables are defined as follows:
\begin{eqnarray}
\tilde{k}^{(*)\mu}(x) & = &
k^{(*)\mu} - k_\nu^{(*)} u^\mu(x) u^\nu(x), \\
\Omega_k(x) & = & \sqrt{m_0^2 - \tilde{k}^{\mu}\tilde{k}_\mu -
\delta M^2(x,~\tilde{k}) }, \\
\tilde{k} &= & \sqrt{-\tilde{k}^{\mu}\tilde{k}_\mu}, \\
k^{\mu}_{\pm i}(x) & =
& u^\mu(x) \omega_{k_i} \mp \tilde{k}_i^{\mu}(x),\\
k^{*\mu}_{\pm i}(x) & =
& u^\mu(x) \Omega_{k_i}(x) \mp \tilde{k}_i^{*\mu}(x),\\
n_{i,j} (x) & = & 1/\!\left[ 
        \exp[ (K^{* \mu}_{i,j}(x) u_{\mu}(x) \! - \! \mu(x)) / T(x) ] 
- 1 \right]\!\! ,\!\! \\
r(i,j,x) & = & \frac{1}{2}\log
        \left[( K^{\mu}_{i,j} u_\mu (x))/ 
        (K^{* \nu}_{i,j}(x) u_\nu(x) ) \right],
        \label{e:rxk}\\
c_{i,j} & =  & \cosh[r({i,j},x)] , \quad 
s_{i,j} \, = \, \sinh[r({i,j},x)],
\end{eqnarray} 
where $i,j = \pm 1,\pm 2$  and the mean and the relative momenta
for the $a$($b$)-quanta are defined as  
$K^{(*) \mu}_{i,j}(x) = [k^{(*) \mu}_i(x) + k^{(*) \mu}_j(x) ]/2 $ and
$q^{(*) \mu}_{i,j} = k^{(*) \mu}_i - k^{(*) \mu}_j , $ respectively.
We assume of course that the mass shift 
and hence squeezing is non-vanishing over only
a finite domain of the freeze-out hypersurface.
In terms of eqs.~(\ref{e:gc},\ref{e:gs}), 
the following new expressions describe the 
particle spectra and the correlation function in the presence of local
squeezing:
\begin{eqnarray}
        N_1({\bf k}_1) & = & G_c(1,1), \\
C_2({\bf k}_1,{\bf k}_2) & =  &
        1 
        + 
        {|G_c(1,2) |^2 \over G_c(1,1) G_c(2,2) }
        + {|G_s(1,2)|^2\over G_c(1,1) G_c(2,2) } .\nonumber
\end{eqnarray}
Fig. 2 illustrates the novel character of 
BBC for two identical bosons caused
by medium mass-modifications, 
along with the familiar Bose-Einstein or HBT correlations on the
diagonal of the $({\bf k}_1 , {\bf k}_2)$ plane.

{\it Particle-antiparticle pairs} --- 
As the Bogoliubov transformation always mixes particles
with antiparticles, the above considerations hold only
for particles that are their 
own antiparticles, e.g. the $\phi$ meson and $\pi^0$.
However, the extension to particle -- antiparticle correlations is 
straightforward. 
Let ``$+$" label particles,
``$-$" antiparticles if antiparticle is different from particle,
let ``$0$" label both particle and antiparticle if they are identical
particles. The non-trivial correlations from mass-modification
for pairs of $(++)$, $(+-)$ and $(00)$ type read as follows:
\begin{eqnarray}
C^{++}_2({\bf k}_1,{\bf k}_2)\! & =  & \!
        1 
        + 
        {|G_c(1,2) |^2 \over G_c(1,1) G_c(2,2) }, \\
C^{+-}_2({\bf k}_1,{\bf k}_2)\! & =  & \!
        1 
        + {|G_s(1,2)|^2\over G_c(1,1) G_c(2,2) } , \\
C^{00}_2({\bf k}_1,{\bf k}_2)\! & =  &\!
        1 
        + 
        {|G_c(1,2) |^2 \over G_c(1,1) G_c(2,2) }
        + {|G_s(1,2)|^2\over G_c(1,1) G_c(2,2) } ,  \nonumber \\
\null & \null & 
\label{e:cfin} 
\end{eqnarray}
where we have assumed for simplicity that mass-modifications of particles 
and antiparticles are the same.

{\it Summary} ---
We formulated the theory of particle correlations and spectra
for bosons with in-medium mass-shifts, which  predicts
the existence of {\it unlimited} back-to-back correlations of 
$\phi\phi$, $K^+K^-$, $\pi^0 \pi^0$ and $\pi^+ \pi^-$ pairs
that  could be searched for at CERN SPS and
upcoming RHIC BNL heavy ion experiments~\cite{lz}.
These correlations not only survive orders of magnitude 
finite time suppressions, but may also be utilized to
determine the  freeze-out time distribution 
and in-medium hadronic mass modifications.  

{\it Acknowledgments}:
We are grateful to T. Bir\'o,
V. N. Gribov, G. Gustafson, P. L\'evai, S. S. Padula 
Yu. M. Sinyukov and R. M. Weiner for stimulating discussions.  
This research was supported by
the DOE's Institute for Nuclear Theory (Seattle), the
US - Hungarian Joint Fund, the Hungarian OTKA grant
T024094, the Hungarian Soros foundation,
the U.S. Department of Energy contracts No. DE-FG02-93ER40764,
DE-FG-02-92-ER40699 and DE-AC02-76CH00016 and the Grant-in-Aid
for Scientific Research No. 10740112 of the Japanese Ministry
of Education, Science and Culture.

\vfill

\null
\vspace*{7.3cm}
\includegraphics{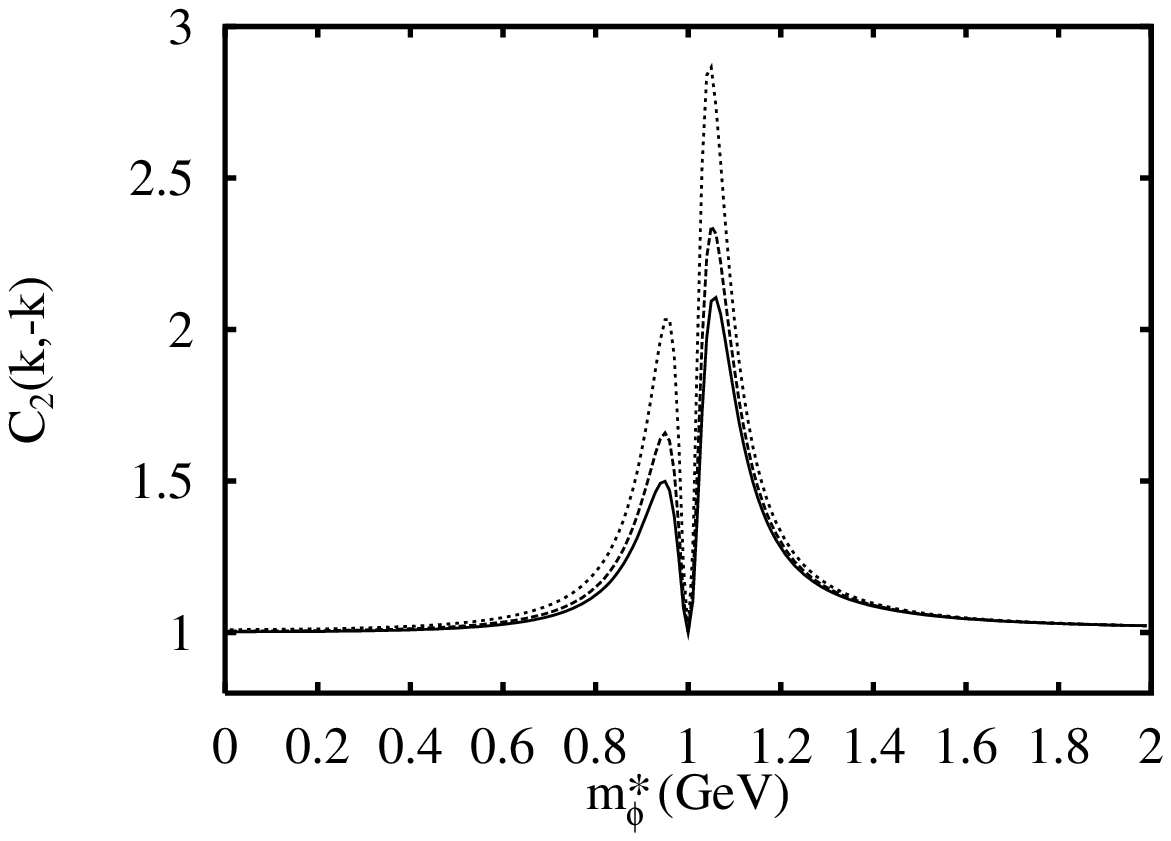}
\vspace{-1.7cm}
\null
        {\bf Fig. 1.} 
	Dependence of the back-to-back correlations (BBC) 
        on the medium modified $\phi$ meson mass, $m^*_{\phi}$,
        for $T = 140$ MeV and $\mu = 0$, 
        where solid, dashed and dotted lines
        stand for $|{\bf k}| = 0 $, 300 and 500 MeV, respectively.
        The magnitude of the BBC is large
        in spite of the finite time  suppression factor
        $1/[1 + (2 \delta t \omega_{\bf k})^2]\sim 0.001$,
        for $\delta t = 2$ fm/c. 

\vfill
\null
\vspace*{7.3cm}
\includegraphics{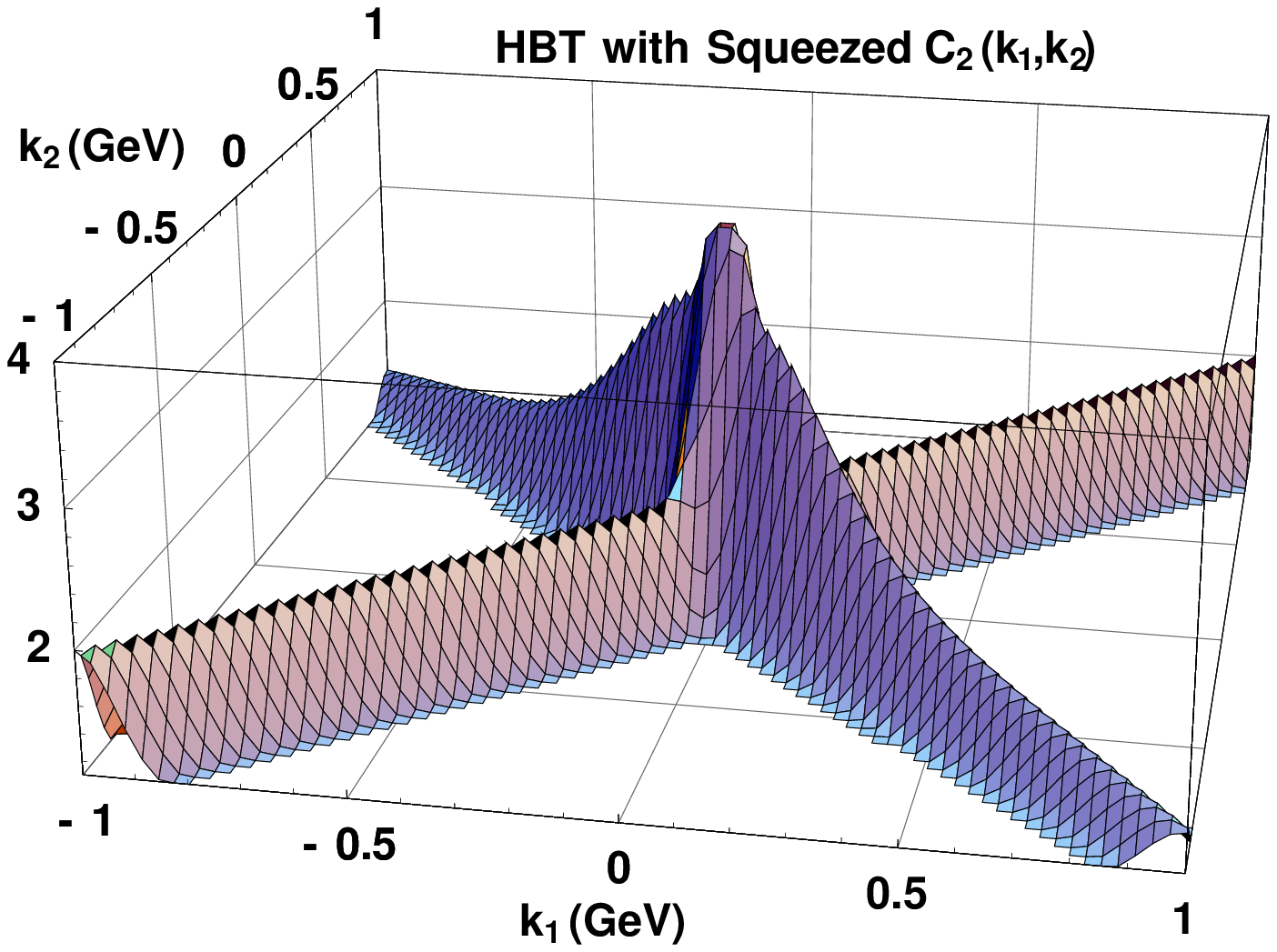}
\vspace{-1.7cm}
\null
       {\bf Fig. 2.}  Schematic illustration of the new kind of correlations
        for mass shifted $\pi^0$ pairs, 
	assuming $T = 140 $ MeV, $G_c \sim \exp[-q_{12}^2 R_G^2/2 ], 
\;G_s \sim \exp[-2 K_{12}^2 R_G^2]$, 
        with $R_G = 2$ fm. 
The fall of the BBC for increasing
values of $ |{\bf k} |$ is controlled here by a momentum-dependent 
effective mass,
$m^*_{\pi} = m_{\pi} [1 + \exp( - {\bf k}^2 / \Lambda_s^2) ] $
with $\Lambda_s = 325 $ MeV in the sudden approximation.
Without the $\Lambda_s$ cutoff, the 
BBC would increase indefinitely 
as $|{\bf k}| \rightarrow \infty $.
\vfill
\end{multicols}
\vfill\eject
\end{document}